\documentclass{webofc}
\usepackage[varg]{txfonts}   

\usepackage{amssymb}

\newcommand{\be}{\begin{equation}}
\newcommand{\ee}{\end{equation}}

\begin{document}

\title{Overlap between Lattice QCD and HRG  with   in-medium effects\\ and parity doubling\thanks{Presented at {\em Strangeness in Quark Matter}, Utrecht, the Netherlands, 10-15 July 2017}
}

\author{
	\firstname{Kenji} \lastname{Morita}\inst{1,2}
\and	
\firstname{Chihiro} \lastname{Sasaki}\inst{1}
\and
\firstname{Pok Man} \lastname{Lo}\inst{1,3}
\and
\firstname{Krzysztof} \lastname{Redlich}\inst{1,2}\fnsep\thanks{Speaker, \email{redlich@ift.uni.wroc.pl}}
}

\institute{
	Institute of Theoretical Physics, University of Wroc{\l}aw, PL-50204, Wroc{\l}aw, Poland
\and
       iTHES Research Group, RIKEN, Saitama 351-0198, Japan
\and
         ExtreMe Matter Institute EMMI, GSI, D-64291 Darmstadt, Germany
          }

\abstract{
  We investigate the fluctuations and correlations involving baryon
 number in hot hadronic matter with modified masses of
 negative-parity baryons,  in the context of the hadron resonance gas. Temperature-dependent masses  are adopted from the recent lattice QCD results and from a chiral
 effective model which implements the parity doubling structure with respect to
 the chiral symmetry. Confronting the baryon number susceptibility,
 baryon-charge correlation, and baryon-strangeness correlation and their
 ratios with the lattice QCD data, we show that the strong downward mass
 shift in hyperons can accidentally reproduce some  correlation ratios, however
 it also tends to overshoot the individual fluctuations and
 correlations.
  This indicates,  that
in order to correctly account for  the influence of the chiral symmetry restoration  on
the fluctuation observables, a consistent framework   of in-medium effects   beyond hadron mass shifts is required.
}

\maketitle


\section{Introduction}
Fluctuations and correlations of conserved charges provide diagnostic
tools for the nature of strongly interacting matter described
by Quantum Chromodynamics (QCD). The first-principles
calculations by lattice QCD (LQCD) have provided not only
equation of state at the physical quark masses
 but also the fluctuations and
correlations of the net-baryon, net-electric charge, and net-strangeness
\cite{Karsch:2017zzw,Bazavov:2017tot,Ding:2015ona,borsanyi10:_qcd,Bazavov2014,bazavov12:_fluct_and_correl_of_net}.
The fluctuations and correlations can also be measured in heavy-ion
collisions to  identify  the state of  created  matter. In particular,
non-Gaussian (higher-order) fluctuations have been expected to probe
critical properties of the system, such as   the QCD critical point in the
beam-energy scan program at RHIC and a remnant of the $O(4)$ criticality at
small baryon density \cite{ejiri06:_hadron_fluct_at_qcd_phase_trans,friman11:_fluct_as_probe_of_qcd}.

For a physical interpretation of the thermodynamic properties, the hadron
resonance gas (HRG) model \cite{Statmodelreview_QGP3,Andronic:2017pug} has been used as
a reference. The equation of state and fluctuations have been well
described by the model  below
the chiral crossover temperature $T = 154\pm 9$ MeV
\cite{karsch11:_probin_freez_out_condit_in,bazavov12:_chiral_and_decon_aspec_of_qcd_trans},
and the experimental data of fluctuations have been analyzed in terms of a gas
of hadronic states
\cite{garg13:_conser,Alba:2014eba,Braun-Munzinger2015,KMR2015}.
At vanishing baryon density, the critical behavior due to the chiral
phase transition  appears at the sixth order of net-baryon fluctuations
\cite{friman11:_fluct_as_probe_of_qcd}. This is because  the singular contribution to the free energy is
suppressed  in the lower-order fluctuations even in
the vicinity of the expected second order transition. Therefore, one expects that the lower order fluctuations
and correlations may be well described by the hadronic degrees of freedom, and
they would provide a reliable baseline for exploring the critical behavior
\cite{ejiri06:_hadron_fluct_at_qcd_phase_trans}.

According to the recent LQCD calculations, however, some  correlations between
conserved charges cannot be explained by a conventional HRG model.
Particularly interesting quantities are those involving net-baryon
number,  baryon-charge (BQ) and baryon-strangeness (BS) correlations.
Since mesons do not contribute to these quantities, one may gain
access to a role of baryonic degrees of freedom and interactions which are not visible in
the equation of state and other meson-dominated quantities due to their
heavy masses \cite{Lo:2017lym}.

In the HRG model the interaction of hadrons is replaced by
 resonances, and in the first approximation their widths can be  neglected. Then, the partition
function of the interacting hadronic system can be written as a mixture of free gases
of all  stable and resonant hadrons \cite{Dashen1969}. The validity of
the vanishing-width approximation for the fluctuations has been recently
examined based on the S-matrix formalism for $K$--$\pi$
\cite{Friman:2015zua} and $\pi$--$N$--$\Delta$ \cite{Lo:2017ldt} systems. It
has been found that an explicit treatment of the width can have
a substantial effect on the fluctuations.

On the other hand, at finite temperature and density one expects changes
in the spectral property of hadrons. In particular, a search for those
modifications because of the partial restoration of the QCD chiral symmetry has been one of
the central subjects in heavy-ion collisions \cite{Hayano:2008vn}.
Since the vacuum masses are used in the HRG model,
the observed agreement of the equation of state in the HRG and
LQCD suggests no need  for a strong mass reduction in the dominant
degrees of freedom such as light mesons. Although  chiral symmetry
predicts substantial medium modifications of low-lying mesons
\cite{Hatsuda93,Hohler2014}, their reliable estimates in
LQCD have been limited to the screening masses \cite{Maezawa:2016pwo}.

Recently,  masses of the non-strange and strange baryons with positive and negative parity
have been extracted from the temporal correlation functions by the FASTSUM  collaboration~\cite{Aarts:2017rrl,Aarts_SQM2017}.
The negative-parity states clearly show downward mass
shifts, whereas the positive-parity baryons stay insensitive to temperature.
The obtained spectra  follow an expectation from the parity doublet
picture of the chiral symmetry, indicating that the masses of a negative- and positive-parity partners tend to  degenerate when approaching the chiral crossover. The above medium
modification was used to possibly  explain a missing contribution in the correlations between the conserved charges
\cite{Aarts_SQM2017}.

In the following  we
investigate the fluctuations and correlations of the net-baryon with
net-charge and net-strangeness on the basis of the  HRG model implementing in-medium mass
modifications. We employ temperature dependent masses of the negative-parity
octet and decouplet baryons from the lattice QCD and from a chiral
effective model with parity doubling \cite{sasaki17:_parity}. We show  that the strong downward mass
 shift in hyperons can accidentally reproduce some  correlation ratios, however
 it also tends to overshoot the individual fluctuation and
 correlation observables. This indicates that  to quantify fluctuation and correlation of conserved charges in the presence of chiral symmetry restoration it is not sufficient to implement in-medium hadron masses in the statistical sum of the hadron resonance gas.


\section{Baryon masses in parity doublet picture}

\subsection{Lattice QCD}

\begin{table*}[!t]
 \caption{Assignment of negative parity states in in-medium HRG}
 \label{tbl:negativeparitybaryon}
 \begin{tabular}[t]{l|cccccccc}\hline
 P$^-$ State & $N$ & $\Lambda$ & $\Sigma$ & $\Xi$ & $\Delta$ & $\Sigma^*$ & $\Xi^*$ & $\Omega$ \\\hline
  Mass { LGT} \cite{Aarts:2017rrl,Aarts_SQM2017} & 1779 & 1899 & 1823 & 1917 & 2138 & 2131 & 2164 & 2193 \\
  $M_-^i(T_c)$ [MeV]& 1254  & 1172  & 1329  & 1295 & 1405 & 1398 & 1426 & 1383 \\
  $b_i$             & 0.338 & 0.369 & 0.257 & 0.275 & 0.312 & 0.257 & 0.246 & 0.213 \\ \hline
  Assignment A & $1535$ & $1405$ & $1750$ & $1690$ & $1700$ &
			  $1670$ & $1820$ & 2250
				  \\
  Assignment B & $1535$ & $1670$ & $1750$ & $1950$ & 1700 & $1940$ & $1820$ & $2250$ \\ \hline
  Mass  [MeV] & 1535 & 1790 & 1880 & 2090 & 1710 & 1930 & 2150 & 2380 \\
  Assignment C & 1535 & $1800$ & $1880$ & $2120$ & $1700$ &
			  $1940$ & $2250$ & $2380$ \\ \hline
 \end{tabular}
\end{table*}

The FASTSUM collaboration presented the
temperature dependent masses of $N$, $\Delta$, and $\Omega$ states,
extracted from imaginary time correlators of the
corresponding interpolating operators in $N_f = 2 + 1$ lattice simulations ~\cite{Aarts:2017rrl}.
Their calculations were performed for heavier light-quark mass than the physical one;
$m_\pi=384$ MeV, while the strange quark is set to the physical one.
Thus, the mass of baryons except $\Omega$ are heavier than the physical
ones. This yields the nucleon with positive parity $N_+$ has
$m_{N_+}=1158$ MeV (939 MeV in PDG), while the positive parity $\Omega_+$
has $m_{\Omega_+}=1661$ MeV (1672 MeV in PDG). The results of other octet
and decuplet states have been shown in \cite{Aarts_SQM2017}. Here we
use the masses at $T/T_c=0.24$, $0.76$, $0.84$, and $0.95$. Note that
owing to the heavier pion mass, the crossover temperature $T_c=185$ MeV,
which was determined from the renormalized Polyakov loop, is also higher than
the physical one, $T_c = 154$ MeV.

\begin{figure}[!h]
 \centering
 \includegraphics[width=0.6\columnwidth]{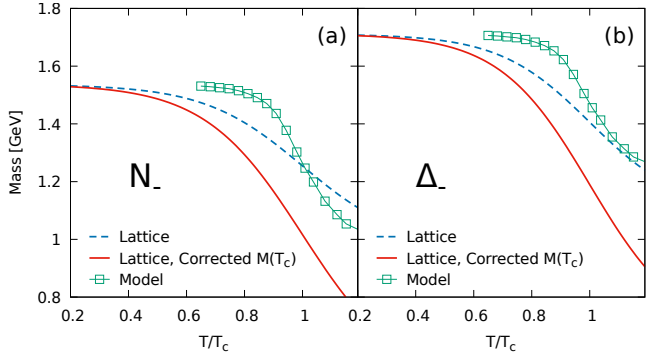}
 \caption{Temperature dependence of the mass of $N_-$ (a) and $\Delta_-$
 (b) from lattice-motivated parameterization Eq.~\eqref{eq:fit} (lines) and
 from a chiral effective model (squares).}
 \label{fig:N_and_Delta_mass}
\end{figure}

In Table \ref{tbl:negativeparitybaryon}, we list the masses of the
octet and decuplet negative-parity states at $T=0.24T_c$.
They are well parameterized, for the given the state $i$, by~\cite{Aarts:2017rrl,Aarts_SQM2017}
\begin{align}
 M_-^i(T) &= M_-^i(T=0)\omega(T,b_i) + M_-^i(T_c)(1-\omega(T,b_i))\label{eq:fit}\\
 \omega(T,b_i)&= \tanh[(1-T/T_c)/b_i] / \tanh(1/b_i)\label{eq:tanh}
\end{align}
where $M_-^i(T=0)$ is fixed to be the value obtained at $T=0.24T_c$ and
$M_-^i(T_c)$ and $b_i$ are the fitting parameters fixed for each
channel. The data are given as a function of $T/T_c$; the value of $T_c$
does not affect the fit but later it is set to the physical value
154~MeV. The parameter $b_i$ corresponds to the width of the chiral
crossover. The results of the fits are shown in Table
\ref{tbl:negativeparitybaryon}.

\begin{figure*}[!tb]
 \centering
 \includegraphics[width=0.7\textwidth]{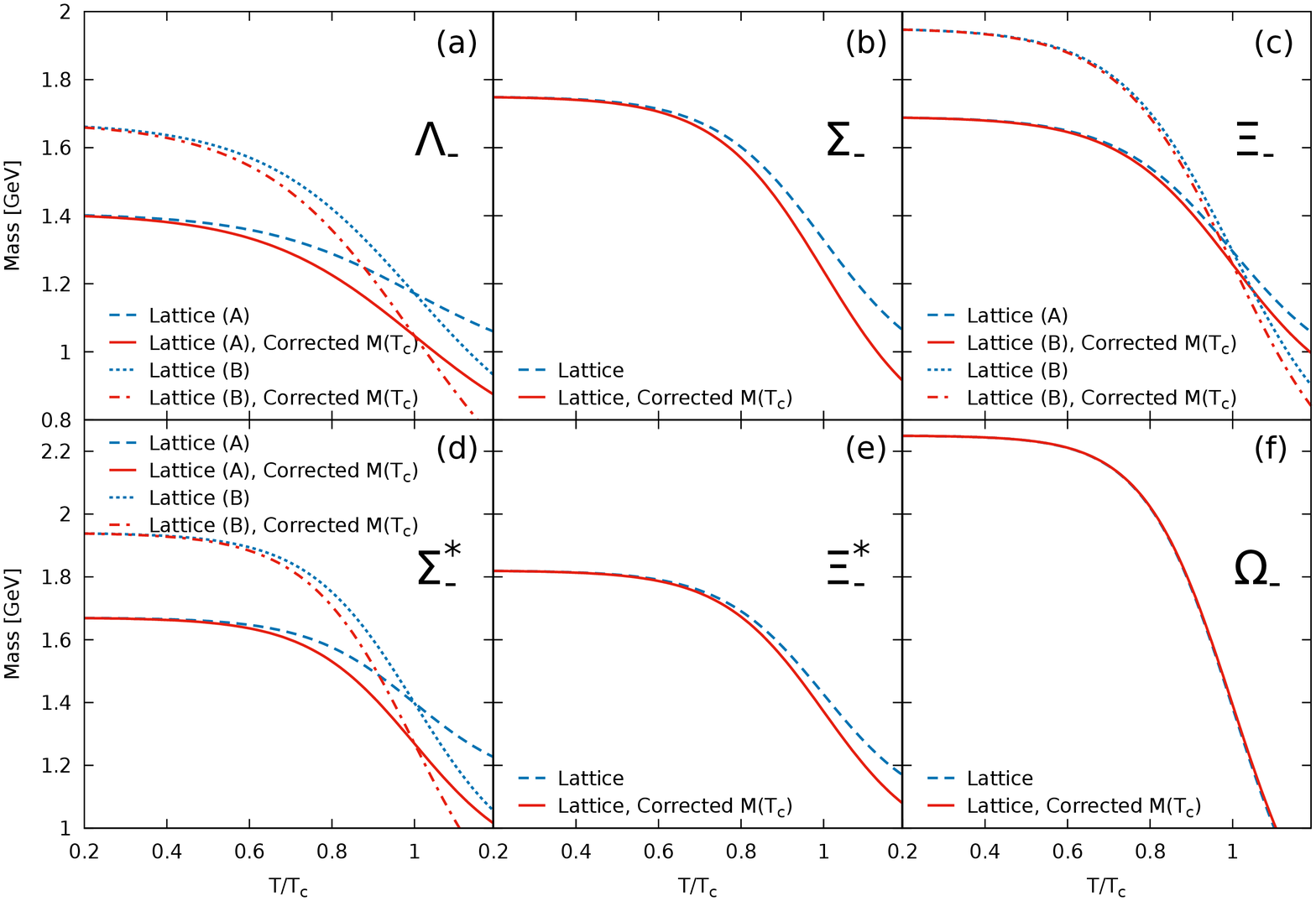}
 \caption{Temperature dependence of octet (upper row) and decuplet
 (lower row) hyperons from lattice-motivated parameterization
 Eq.~\eqref{eq:fit}. }
 \label{fig:hyperonAB}
\end{figure*}

To correct the unphysical effect from the heavy up and down quarks, $M_-^i(T=0)$ is set to its PDG
mass in the following. The value of $M_-(T_c)$ needs to be corrected as well,
and we shall re-scale it by multiplying the
factor $M_+^{\text{PDG}}(T=0)/M_+^\text{lattice}(T=0)$ with $M_-(T_c)$.

It is not clear how to assign the negative-parity states to the observed ones
because of unknown quantum numbers of some of the  candidates.
We follow the suggestion in \cite{Aarts_SQM2017} and
further adopt two different assignments, A and B.
The set A takes the lighter state for $\Lambda_-$, $\Xi_-$, and $\Sigma^*_-$
while the set B does the heavier ones.

The obtained temperature-dependent masses are displayed in
Figs.~\ref{fig:N_and_Delta_mass} and \ref{fig:hyperonAB}. The masses
of negative-parity states start to drop around $T/T_c \simeq 0.6$, then approach those of
their positive-parity partners. In comparing the non-strange baryons with the hyperons, one
finds the correction of $M(T_c)$ significantly affects the baryons with more
light quarks, as expected.

\subsection{Chiral model}

The parity-doubled baryons can be modeled in an effective chiral approach~\cite{Detar:1988kn},
where a chiral-invariant mass is naturally introduced.
In \cite{sasaki17:_parity}, the mass relations to the light quark $\sigma_q$ and
strange quark $\sigma_s$ condensates are given for the octet and decuplet states as~\footnote{
 In~\cite{sasaki17:_parity} the same $m_0$ was assumed common to the octet and decuplet
 baryons. Here we lift this constraint and introduce two independent masses. Their
 splitting can be deduced from the spin-spin interaction.
}
\begin{eqnarray}
m_{N_\pm}&=&
\left(a_N \mp b_N\right)3\sigma_q + m_{N0}\,, {\hskip 1.6cm~~~~~~~~~~m_{\Delta_\pm}
=
\left(a_\Delta \mp b_\Delta\right)3\sigma_q + m_{\Delta 0}\,}
\nonumber\\
m_{\Sigma_\pm}
&=&
\left(a_N \mp b_N\right)\left(2\sigma_q + \sqrt{2}\sigma_s\right)
{}+ m_{N0} + m_1\,, ~~m_{\Sigma^\ast_\pm}
=
\left(a_\Delta \mp b_\Delta\right)\left(2\sigma_q + \sqrt{2}\sigma_s\right)
{}+ m_{\Delta 0} + m_s\,
\nonumber\\
m_{\Lambda_\pm}
&=&
\left(a_N \mp b_N\right)\left(2\sigma_q + \sqrt{2}\sigma_s\right)
{}+ m_{N0} + m_3\,, ~~ m_{\Xi^\ast_\pm}=
\left(a_\Delta \mp b_\Delta\right)\left(\sigma_q + 2\sqrt{2}\sigma_s\right)
{}+ m_{\Delta 0} + 2m_s\,
\nonumber\\
m_{\Xi_\pm}
&=&
\left(a_N \mp b_N\right)\left(\sigma_q + 2\sqrt{2}\sigma_s\right)
{}+ m_{N0} + m_2\,, ~~m_{\Omega_\pm}
=
\left(a_\Delta \mp b_\Delta\right)3\sqrt{2}\sigma_s
{}+ m_{\Delta 0} + 3m_s\,.
\end{eqnarray}
The parameters $m_{1,2,3}$ are related via the Gell-Mann--Okubo relation as
\begin{equation}
m_1 = 2m_2 - 3m_3\,.
\end{equation}
The parameters in the above expressions are determined at zero temperature
as in Table~\ref{paraB}.
\begin{table*}
\begin{center}
\begin{tabular*}{12cm}{@{\extracolsep{\fill}}ccccccc}
\hline
$a_N$ & $b_N$ & $m_1$ [GeV] & $m_2$ [GeV] &
$a_\Delta$ & $b_\Delta$ & $m_s$ [GeV]
\\
\hline
$1.22$ & $1.08$ & $0.248$ & $0.367$ &
$1.16$ & $0.862$ & $0.139$
\\
\hline
\end{tabular*}
\end{center}
\caption{
Set of parameters in the baryon-mass relations. We set $m_{N0} = 0.9$ GeV
and $m_{\Delta 0} = 1.15$ GeV.
}
\label{paraB}
\end{table*}

The masses at $T=0$ are also summarized in
Table \ref{tbl:negativeparitybaryon}.
Thermal modifications of the baryon masses are driven by the quark condensates.
Following \cite{sasaki17:_parity}, we shall use the in-medium condensates measured
in LQCD by the HotQCD collaboration~\cite{bazavov12:_chiral_and_decon_aspec_of_qcd_trans}.
One readily finds that some of the hyperon states have larger masses than those
expected from the lattice baryon-spectrum in~\cite{Aarts:2017rrl,Aarts_SQM2017}.
Thus, we assign these hyperons to the PDG
states as in the bottom line of Table \ref{tbl:negativeparitybaryon}
and refer to set C. For a comparison with LQCD, we also apply the set C
to the lattice-motivated parametrization \eqref{eq:fit}.

The non-strange baryons exhibit the thermal behavior as in Fig.~\ref{fig:N_and_Delta_mass}.
To some extent, their trends are similar to the results by the FASTSUM collaboration,
whereas the parametrization~(\ref{eq:tanh}) leads to a sizable difference in the mass
dropping near $T_c$. This is particularly evident when the re-scaling of $M_-(T_c)$ is made.

\begin{figure*}[!t]
 \centering
 \includegraphics[width=0.7\textwidth]{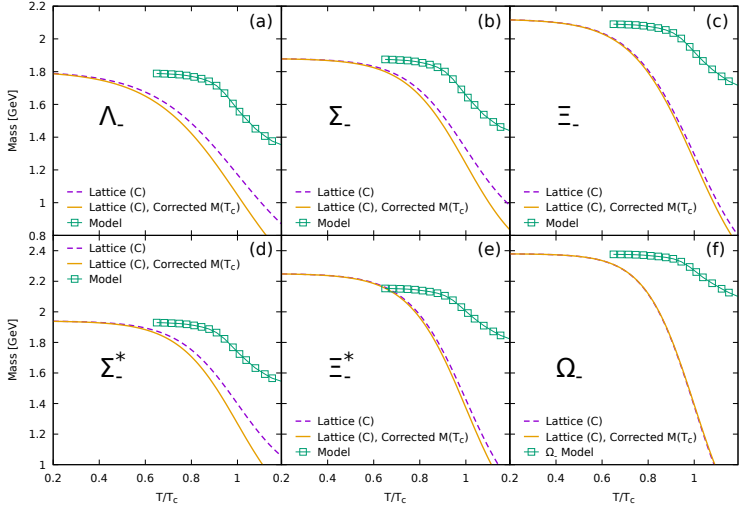}
 \caption{Comparison of hyperon masses between the lattice-motivated
 parametrization (lines) and the chiral effective model (squares) in the
 assignment C. As in Fig.~\ref{fig:hyperonAB}, the  octet and decuplet
 hyperons are shown in upper and lower panels, respectively. }
 \label{fig:hyperonC}
\end{figure*}

Figure \ref{fig:hyperonC} displays the mass of the hyperons in the
assignment C.
The most distinct
difference from the non-strange baryons lies in the mass of the hyperons near $T_c$; the model
calculation shows a rather moderate downward-shift, while the lattice
QCD results lead to a strong mass drop. Comparing $\Lambda$ ($\Sigma$),
$\Xi$, and $\Omega$ states between the model and the lattice-inspired scaling,
one clearly finds that the difference becomes more
significant as the baryon contains more strangeness. This
may be attributed to the difference in the masses between the light and strange quarks
; The FASTSUM setup with a large pion mass $m_\pi=384$ MeV and the physical kaon mass
is rather close to the flavor SU$_f$(3) limit, while the model calculation done with the physical
pion mass is dominated by SU$_f$(2). Therefore, it will be  intriguing
to see whether the strong mass reduction of $\Omega$ and $\Xi$ would still persist
in the lattice simulations with a lighter pion.

\section{Fluctuations and correlations from in-medium HRG}
\label{sec:HRG}

We employ the hadron resonance gas model to explore the fluctuations and
correlations in the hot hadronic matter. In this work we entirely
rely on the zero-width approximation; all the resonances are treated as
point-like particles and their spectral functions are given by the delta
function. Consequently, the thermodynamic pressure at temperature $T$ and
chemical potentials ${\mu_j}$ $(j=B,Q,S)$ is given by the summation of
ideal gas pressure of each particle species as
\begin{equation}
 p^{\text{HRG}}(T,\mu_B,\mu_Q,\mu_S) = \sum_{i=\text{hadrons}}p^{\text{ideal}}(T,\mu_i;m_i)
\end{equation}
where $\mu_i = B_i \mu_B + Q_i \mu_Q + S_i \mu_S$ is the chemical
potential of particle $i$ with baryon number $B_i$, electric charge
$Q_i$, and strangeness $S_i$ and the ideal gas pressure with degeneracy $d$ reads
\begin{equation}
 p^{\text{ideal}}(T,\mu;m) = \mp \frac{d T}{2\pi^2} \int_{0}^{\infty}\!\! dp
  p^2 \ln[1 \mp e^{-(\sqrt{p^2+m^2} - \mu)/T}].
\end{equation}
The signs are negative for mesons and positive for baryons.
The fluctuations and the correlations at vanishing chemical potentials
can be expressed as the generalized susceptibilities
\begin{equation}
 \chi_{ijk}^{BQS} \equiv \frac{\partial^{i+j+k}p^{\text{HRG}}/T^4}{\partial^i(\mu_B/T)
  \partial^j(\mu_Q/T) \partial^k(\mu_S/T)}
\end{equation}

Although
heavy particles ($m \gg T$) are thermally suppressed, the existence of
missing states are expected likely in the strange sectors
\cite{PhysRevLett.113.072001,Lo2015}. In fact, the assignment C in
Table~\ref{tbl:negativeparitybaryon} contains several 1-star states
whose spins are undetermined in the latest PDG \cite{pdg2016}.
For those state we shall assume the same spin with its partner. For
consistency we also include other unconfirmed state; we have in total 28
nucleons, 22 $\Delta$ baryons, 18 $\Lambda$ baryons, 21 $\Sigma$
baryons, 8 $\Xi$ baryons, 3 $\Omega$ baryons, and their isospin
multiplets. We also include a deuteron, triton, $^3$He and $^4$He as they
are considered to be thermal ingredients in QCD thermodynamics and heavy
ion collisions \cite{Andronic:2017pug}.

In this work we focus on $\chi^{B}_2$, $\chi^{BQ}_{11}$, and
$\chi^{BS}_{11}$ at $\mu_B=\mu_Q=\mu_S=0$.
Since $\chi_{ijk}^{BQS} \propto (B/T)^{i}(Q/T)^j (S/T)^k$,
mesons do not contribute to these susceptibilities. Thus they
are good measures of in-medium effects in the baryon sector on the QCD
thermodynamics. In general, particle masses in a medium
can depend not only on temperature but also on chemical potentials. We neglect
such an intrinsic chemical-potential dependence in the baryon masses in
the present calculations. This is well justified at small chemical potential
since the quark condensates are not much affected.

\begin{figure*}[!t]
 \centering
 \includegraphics[width=0.99\textwidth]{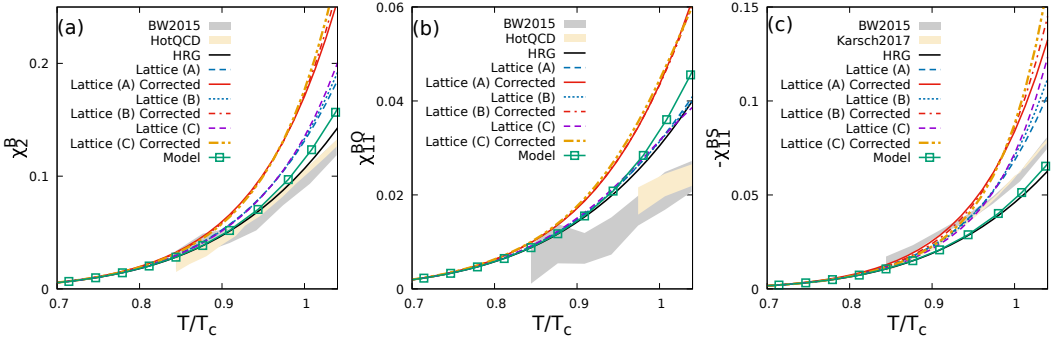}
 \caption{Fluctuations and correlations from the HRG model with and
 without the mass shifts. From left to right, (a) Baryon number susceptibility, (b)
 Baryon-charge correlation, and (c) Baryon-strangeness correlation are
 shown together with the LQCD results from HotQCD
 \cite{bazavov12:_fluct_and_correl_of_net,karsch:QM17} and
 Budapest-Wuppertal Collaboration \cite{Bellwied:2015lba}}
 \label{fig:chiB}
\end{figure*}

We display $\chi_2^B$, $\chi_{11}^{BQ}$, and $-\chi_{11}^{BS}$ in
Fig.~\ref{fig:chiB}. In the baryon number susceptibility $\chi_2^B$, the
HRG model agrees with the LQCD results below $T_c$. \footnote{The
present HRG leads to a slightly larger $\chi_2^B$ than the conventional one
because of including the 1-star states and the multi-baryons.}
With the in-medium mass shift, a decrease in the negative-parity baryon masses
lead to an enhancement of the susceptibility. One sees a moderate enhancement in all
the three scenarios, Lattice (A) (B) and (C),  and it becomes stronger when $M(T_c)$ is corrected,
since the correction further reduces the masses just below $T_c$.
On the other hand, the different assignments of the parity-partners lead
to a minor change. The results with the masses from the chiral
effective model follows the same trend, but the amount of the
enhancement is much smaller than the $\tanh$-parameterization, owing to the weaker downward
mass shifts in the hyperon sectors (see Fig.~\ref{fig:hyperonC}.)
In any case, all the results with mass reduction overshoot the lattice
data of $\chi_2^B$.

The $\chi_{11}^{BQ}$ shows a different tendency in Fig.~\ref{fig:chiB}(b); the results of Lattice
(A)-(C) without the correction of $M(T_c)$ do not exhibit any enhancement
from the HRG, even below the ``Model'' result.
In the $\chi^{BQ}_{11}$, $I=1$ states do not contribute because neutral particles do not
contribute and the positively-charged states cancel with the negatively-charged states.
Charged $N$ states and doubly charged $\Delta^{++}$ states
contribute positively, while negatively-charged $\Xi_-^-$, $\Xi_-^{*-}$
and $\Omega_-^-$ states contribute oppositely. Without a strong mass-shift
of these multi-strange states, the $\chi^{BQ}_{11}$ becomes larger due to the
smallness of the hyperon contribution suppressed by the Boltzmann factor, as in the
``Model'' case. With the strong mass shift of the hyperons, they tend to
cancel a part of the positive contribution, and suppress the
$\chi_{11}^{BQ}$.  The strong enhancement with the corrected $M(T_c)$
clearly signals that the effect of the correction is smaller for $\Xi$
states and larger for $N$ and $\Delta$ states, as seen in
Figs.~\ref{fig:N_and_Delta_mass}-\ref{fig:hyperonC}.

The behavior of $-\chi_{11}^{BS}$ in Fig.~\ref{fig:chiB}(c) can be
understood in a similar way to the $\chi_2^B$; the ``Model'' result
does not differ much from the HRG, because of a weaker
mass reduction. Owing to the different trend in the hyperon sectors, the
 curves deviate more sensitively depending on the fate of the masses.

\begin{figure*}[!t]
 \centering
 \includegraphics[width=0.4\columnwidth]{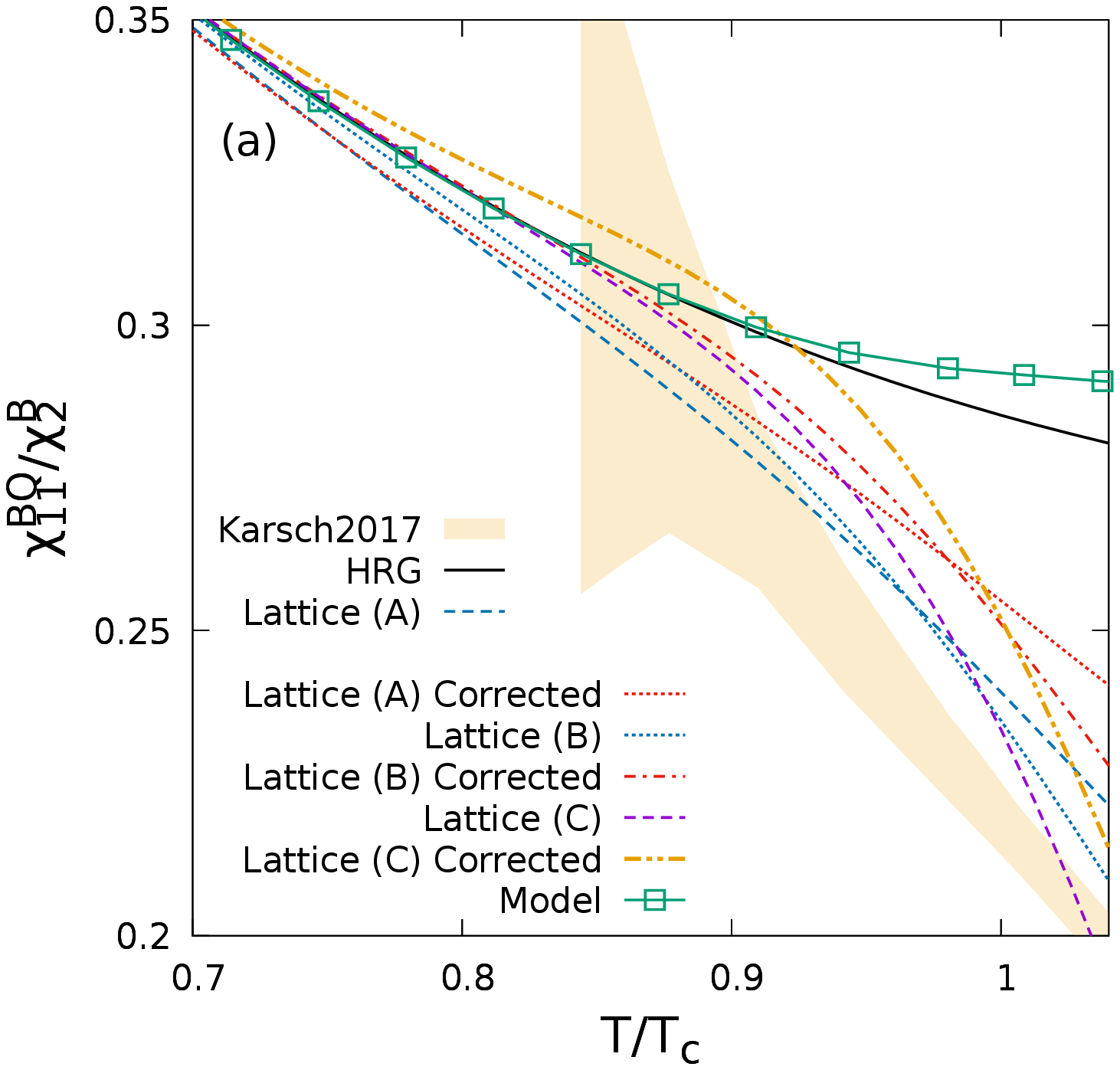}
 \includegraphics[width=0.4\columnwidth]{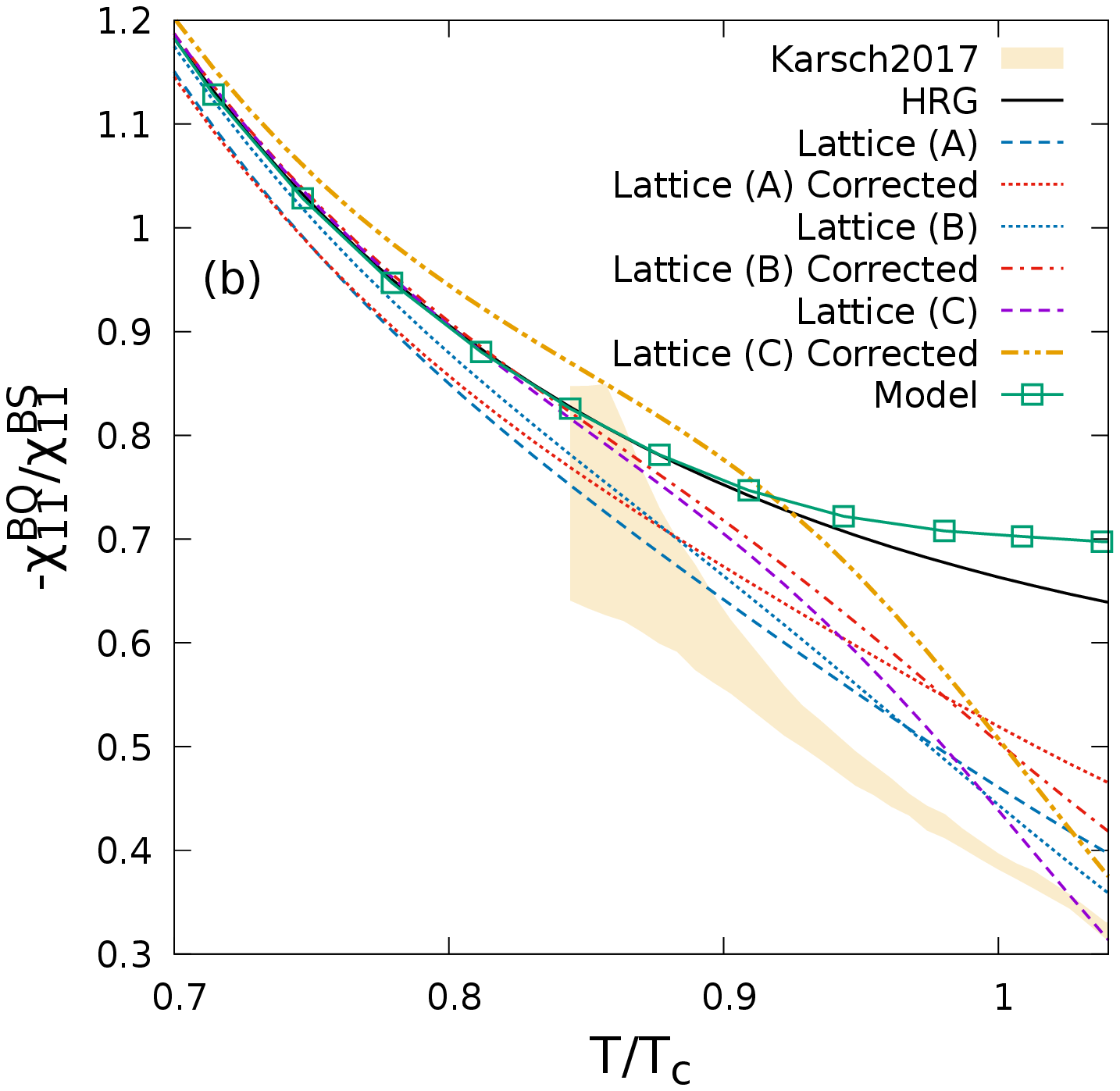}
 \caption{Ratio of the baryon-charge correlation to the baryon number
 fluctuation (a) and the baryon-strangeness correlation (b). The legends
 are the same as Fig.~\ref{fig:chiB}.}
 \label{fig:ratio}
\end{figure*}

The observed differences can be more pronounced by taking the ratio
between the susceptibilities. In Fig.~\ref{fig:ratio}(a), we display the
$\chi_{11}^{BQ}/\chi_2^B$ together with the LQCD data.
The results from Lattice (A) and (B) reproduce the one shown in
\cite{Aarts_SQM2017}; they follows the trend seen in the HotQCD data. As explained
above, however, the individual susceptibilities cannot be reproduced. The
coincidence comes from the reduction of $\chi_{11}^{BQ}$ owing to the
strong mass reductions in the charge asymmetric states. The ``Model''
case shows the opposite trend that the ratio is above the HRG, due to
the much milder mass-reduction in the hyperon sectors.
Therefore, the data cannot be explained solely by the mass
reduction.

Also shown in Fig.~\ref{fig:ratio}(b) is the
$-\chi_{11}^{BQ}/\chi_{11}^{BS}$.
Again, while the sets (A) and (B) follow the trend of the HotQCD data,
the ``Model'' shows the opposite.
This supports the observation that naive insertion of the in-medium
masses of parity doublers into the HRG  results in fluctuation observables which  are inconsistent with LGT data.
Nevertheless, our results do not rule out the manifestation of
parity doublers. Several studies \cite{Friman:2015zua,Lo:2017ldt} have
revealed that the conventional treatment of the particles in the HRG is not
sufficient and a proper inclusion of the width improves the thermodynamics.
Although those analysis
have been limited to the $S-$matrix approaches where interactions
are incorporated via the two-body scattering phase shift in the vacuum,
a consistent treatment of the resonance widths at finite temperature and densities
should  clarify the consequences of the partial restoration of the
chiral symmetry in the baryonic sector of correlations and fluctuations.

\section{Concluding remarks}
\label{sec:conclusion}

We have studied the fluctuations and correlations of conserved
charges by making use of the hadron resonance gas model with in-medium
masses of the low-lying parity-doubled baryons. Motivated by the recent LQCD result on the baryon
masses at finite temperature, we have examined the consequences of several scenarios with in-medium
masses for the negative-parity octet and decouplet baryons,
adopting a lattice-motivated parametrization
and that from a chiral effective theory.
We have
computed the baryon number fluctuation $\chi_2^B$, baryon-charge
correlation $\chi_{11}^{BQ}$, and baryon-strangeness correlation
$\chi_{11}^{BS}$. The results have been understood on the basis of the
presence or absence of the strong mass reduction in the hyperon sectors.
We have pointed out that the reproduction of the LQCD results
of the ratios $\chi_{11}^{BQ}/\chi_2^B$ and
$-\chi_{11}^{BQ}/\chi_{11}^{BS}$ is just accidental when the in-medium
masses are naively introduced in the conventional HRG approach.
Since the strong mass shifts of the hyperons observed in LQCD can be regarded
as a consequence of the approximate SU$_f$(3) nature due to a heavy pion mass,
it is a critical issue to be examined
whether the hyperon mass-shift persists for the physical pion mass.

Although 
our results clarify the interplay of the different particle
species in the behavior of the baryon-charge and baryon-strangeness
correlations, we emphasize that the treatment of the resonances in
the conventional HRG model is insufficient  to quantify  correlation and fluctuations of conserved charges.  This indicates,  that
in order to correctly account for  the influence of the chiral symmetry restoration  on
the fluctuation observables, a consistent framework   of in-medium effects   beyond hadron mass shifts is required.
\subsection*{Acknowledgments}
We acknowledge stimulating discussions with G\'{a}bor Alm\'{a}si and Bengt Friman.
This work has been partly supported
by the Polish Science Center  NCN  under
Maestro grant 2013/10/A/ST2/00106 and RIKEN iTHES project.

\end{document}